\documentclass[mathleft
]{an}
\usepackage{graphicx}
\usepackage{times}
\begin{document}

\Pagespan{1}{}
\Yearpublication{2009}%
\Yearsubmission{2009}%
\Month{2}%
\Volume{330}%
\Issue{5}%
 \DOI{10.1002/asna.200911207}%

\title{Photometric study of the OB star clusters NGC~1502 and NGC~2169 and mass estimation of their members at the University Observatory Jena\thanks{Based on observations obtained with telescopes of the University
Observatory Jena, which is operated by the Astrophysical Institute of
the Friedrich-Schiller-University.}}

\author{M.M. Hohle\inst{1,2}\fnmsep\thanks{Corresponding author:
  {mhohle@astro.uni-jena.de}}
\and  T. Eisenbeiss\inst{1}
\and  M. Mugrauer\inst{1}
\and  F. Freistetter\inst{1}
\and  M. Moualla\inst{1}
\and  R. Neuh\"auser\inst{1}
\and  S. R\"atz\inst{1}
\and  T.O.B. Schmidt\inst{1}
\and  N. Tetzlaff\inst{1}
\and  M. Va\v{n}ko\inst{1}
}
\titlerunning{Photometric study of NGC~1502 and NGC~2169}
\authorrunning{M.M. Hohle et al.}
\institute{
Astrophysikalisches Institut und Universit\"ats-Sternwarte Jena, Schillerg\"asschen 2-3, 07745 Jena, Germany 
\and 
Max-Planck-Institut f\"ur extraterrestrische Physik, Giessenbachstrasse, 85741 Garching, Germany
}

\received{2009 Feb 21}
\accepted{2009 Apr}
\publonline{2009}


\abstract{In this work we present detailed photometric results of the trapezium like galactic nearby OB clusters NGC~1502 and NGC~2169 carried out at the University Observatory Jena. We determined absolute $BV\!RI$ magnitudes of the mostly resolved components using Landolt standard stars. This multi colour photometry enables us to estimate spectral type and absorption as well as the masses of the components, which were not available for most of the cluster members in the literature so far, using models of stellar evolution. Furthermore, we investigated the optical spectrum of the components ADS~2984A and SZ~Cam of the sextuple system in NGC~1502. Our spectra clearly confirm the multiplicity of these components, which is the first investigation of this kind at the University Observatory Jena. 
}

\maketitle


\section{Introduction}

\begin{figure}[t]
  \centering
   \resizebox{\hsize}{!}
{
   \includegraphics[viewport=0 0 320 320, width=0.25\textwidth]{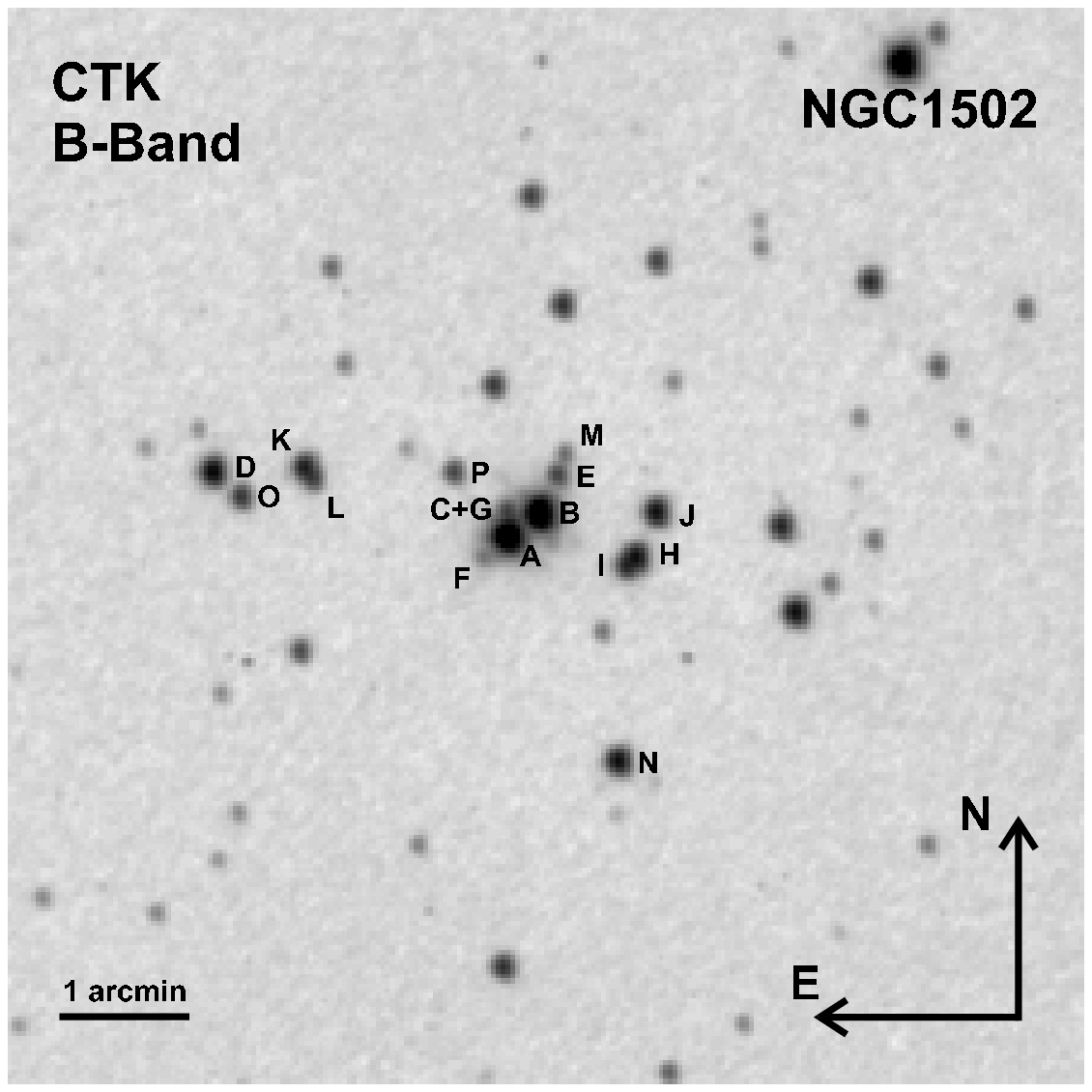}
}
   \caption{The OB star cluster NGC~1502 observed with CTK in the $B$ filter. All 30 $B$-band images from the first observing night are averaged to the image shown here. The designation of the components follows the CCDM catalogue.}
              \label{nsv}
    \end{figure}

The CCDM catalogue ({C}atalogue of the {C}omponents of the {D}ouble
and {M}ultiple stars, Dommanget \& Nys 2000) lists sixteen, not necessarily
physical bound, components of the OB star cluster NGC~1502 (see Fig.\,\ref{nsv}).
The two brightest members with almost equal magnitudes, ADS~2984A and SZ~Cam,
were first noticed as visual double stars by Struve in 1830. Both stars are
spectroscopic binaries themselves (see Plaskett 1924; Morgan, Code \& Withford
1955; Budding 1975) with periods of a few days orbiting a third body, which
is also a (single lined) spectroscopic binary (Lorenz, Mayer \& Drechsel 1998)
with a long term period between 50 and 60 years. This rare configuration of a
hierarchical sextuple system is interesting regarding to star forming processes
(Zinnecker \& Yorke 2007). While Hipparcos lists a parallax value with a large
error of 0.48\,$\pm$\,7.69~mas, i.e. 2100~pc, Lorenz, Mayer \& Drechsel (1998)
obtained 1050~pc from speckle interferometric results using the radial velocity
curves and 836--870~pc from photometric distance estimation for this sextuple system, whereas Chambliss (1992) gives 600~pc. Guthnick \& Prager (1930) first discovered the variable behaviour of SZ~Cam, which is investigated by Lorenz, Mayer \& Drechsel (1998) in $U\!BV$. 

\begin{figure}[t]
  \centering
   \resizebox{\hsize}{!}
{
   \includegraphics[viewport=0 0 320 320, width=0.25\textwidth]{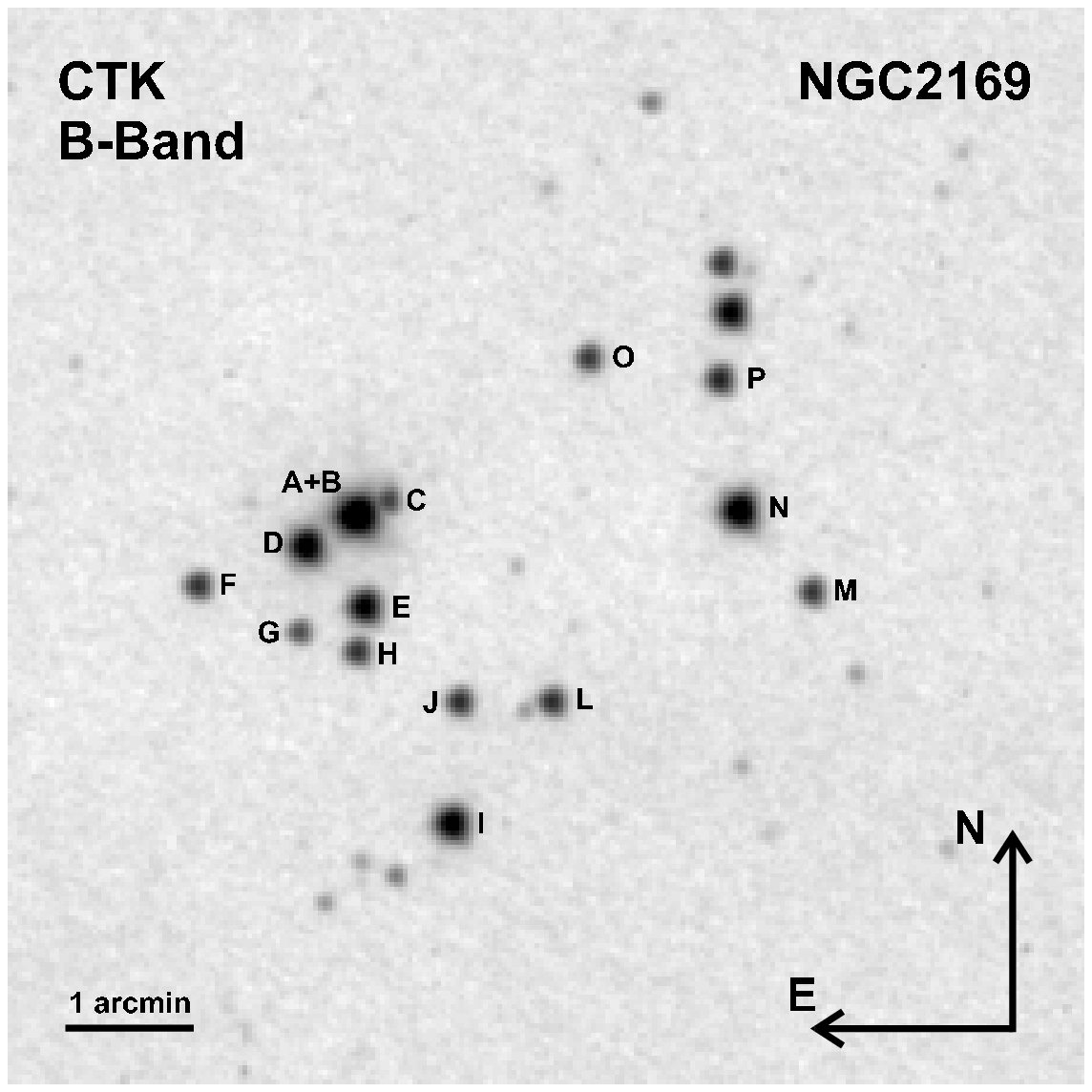}
}
   \caption{Same as in Fig. \ref{nsv} (with CTK in the $B$ filter), but showing the OB star cluster NGC~2169. The K component is too faint to be detected in the individual images.}
              \label{hip}
    \end{figure}

In contrast to this well investigated cluster only a few parameters of NGC~2169 are available. This cluster also hosts sixteen members (CCDM, see Fig.~\ref{hip}) with visual magnitudes ${V\sim7}$ to 10~mag and seems to be quite similar to NGC~1502 and we thus expect a comparable multiplicity. The distance of this cluster given in the Hipparcos catalogue corresponds to 376~pc (2.66\,$\pm$\,1.60~mas), while Echevarr\'{i}a, Roth \& Warman (1979) list 639~pc. They investigated $U\!BV\!RI$ photometry of the members A+B, C, D and E.

We present $BV\!RI$ photometry obtained with the 
Cassegrain-Telescop-Kamera (CTK, see Mugrauer 2009 for a detailed description of this instrument) of \textit{all} sixteen members listed in the CCDM catalogue of both clusters together with estimations of the absorption in the visible band ($A_{V}$) caused by the interstellar medium, spectral type and mass as well as spectra of the spectroscopic binaries ADS~2984A and SZ~Cam done with the FIASCO (Mugrauer \& Avila 2009) spectrograph at the University Observatory Jena.   

This work is the first step of further detailed investigations of trapezium like galactic clusters and is part of a search strategy for radio-quiet, nearby neutron stars described in Hohle, Neuh\"auser \& Tetzlaff (2008). Most of the stars in these clusters are likely massive supernova progenitors in a relatively small volume.

\section{Absolute photometry of NGC~1502 and NGC~2169}

We took 90 images of NGC~2169 (30 each in $BRI$ with twelve, four and five seconds exposure time for the different filters) in the first night (January 8) and 118 in the second (January 9) with ten seconds exposure for 28 images in the $B$ band and 30 images in $V\!RI$ with three seconds exposure time each. For NGC~1502 we took 116 images in $BV\!RI$ (26 in $B$ with twelve seconds exposure time, 30 images in $V\!RI$ with two seconds exposure each) in the first night and 119 in the second (30 in $B$ with six seconds exposure time, 29 in the $V$ and 30 in the $R$ band with two seconds each and 30 images in $I$ with 1.5~s) for absolute photometry.

All images are flat fielded and dark subtracted. In some images illumination effects due to the bright moon are visible. Therefore we applied a subtractive illumination correction using the source extractor (SE, see Bertin \& Arnout 1992), which uses a mesh based background subtraction with a mesh size of 64\,pixels. The objects are detected by thresholding with reference to that mesh based background. This background is finally removed from all images. We determined instrumental magnitudes
for all sources detected in the individual CTK images in all filters using aperture photometry in {ESO-MIDAS} (Banse et al. 1987);
the aperture radius used in our image
for the source photons is 9.8$''$,
the background was measured in a ring
around the source with radii
between 32.5$''$ and 47.7$''$. 

For absolute photometry we observed the Landolt field SA\,96 (Landolt 1992;
Galad\'i-Enr\'iquez et al. 2000) four times in the first night and three times
in the second night (see Table \ref{standardob}). We accomplished the magnitude
calibration with the stars SA\,96\,405 and SA\,96\,406 in this field. The
difference between Landolt magnitudes ($m_{\rm L}$) and $m_{\rm inst}$ defines the
detector zero point $c$ and the atmospheric extinction coefficient $k$ (see
Table \ref{standard}) following 
\begin{equation}
m_{\rm L}-m_{\rm inst}=-k\!\cdot\! Y +c,
\end{equation}
where $Y$ is the airmass. In order to reduce the effect of extinction due to changing airmass 
(${k\!\cdot\! Y}$) we always determined the calibrated magnitudes of all detected objects in the individual CTK images. Except the K component in NGC~2169 (see Fig. \ref{hip}), all objects have a sufficient amount of counts.
The determined values for the different filters are given in Table \ref{standard}.

\begin{table}[t]
\caption{Observations of the Landolt field SA\,96 at the University Observatory Jena for absolute photometry done for four different air masses in the first night and three air masses in the second night with the CTK.}
\label{standardob}
\begin{tabular}{cccccc}
\hline\noalign{\smallskip}
\multicolumn{3}{c}{2009 January 8} & \multicolumn{3}{c}{2009 January 9}\\
Images&Filter&Exp. [s]\phantom{\quad\ \ }&Images&Filter&Exp. [s]\\[1.5pt]
\hline\noalign{\smallskip}
2&$B$&90&2&$B$&90\\
2&$V$&25&2&$V$&25\\
2&$R$&20&2&$R$&20\\
2&$I$&25&2&$I$&25\\
\hline\noalign{\smallskip}
3&$B$&90&2&$B$&70\\
4&$V$&25&2&$V$&20\\
2&$R$&20&2&$R$&15\\
2&$I$&25&2&$I$&20\\
\hline\noalign{\smallskip}
2&$B$&90&3&$B$&70\\
2&$V$&25&2&$V$&20\\
2&$R$&20&2&$R$&15\\
2&$I$&25&2&$I$&20\\
\hline\noalign{\smallskip}
2&$B$&90& & & \\
2&$V$&25& & & \\
2&$R$&20& & & \\
2&$I$&25& & & \\

\hline
\end{tabular}
\end{table}

\begin{table}[t]
\caption{Derived zero points $c$ (normalized for 1s exposure) and atmospheric extinction coefficients $k$ for $BV\!RI$ filters measured during the two observation nights at the University Observatory Jena with the CTK.}
\label{standard}
\begin{tabular}{c c@{\,$\pm\,$}c c@{\,$\pm$\,}c }
\hline\noalign{\smallskip}
Filter & \multicolumn{2}{c}{Zero Point $c$} & \multicolumn{2}{c}{Extinction $k$}\\[1.5pt]
\hline\noalign{\smallskip}
\multicolumn{5}{c}{\qquad\qquad 2009 January 8}\\[1.5pt]
$B$&17.805&0.023&0.289&0.012\\
$V$&18.776&0.015&0.152&0.008\\
$R$&18.790&0.016&0.104&0.008\\
$I$&18.251&0.020&0.058&0.010\\
\hline\noalign{\smallskip}
\multicolumn{5}{c}{\qquad\qquad 2009 January 9}\\[1.5pt]
$B$&17.889&0.011&0.285&0.006\\
$V$&18.874&0.017&0.164&0.009\\
$R$&18.833&0.020&0.110&0.010\\
$I$&18.292&0.020&0.080&0.011\\
\hline
\end{tabular}
\end{table}

For NGC\,1502 the central region of the cluster is barely resolved, thus the PSFs of these objects are merged and the light of neighbouring objects contaminates in the individual apertures. These measurements are excluded from the analysis.

While for NGC~2169 only the A and B components are not resolved (hereafter A+B) and the K component is to faint, for NGC~1502 the C and G (C+G), H and I (H+I) and K and L (K+L) components could not be resolved. The unresolved objects will be treated as one object.

\section{Spectra of ADS~2984A and SZ~Cam}

In addition to the photometric measurements of all components of the cluster NGC~1502 we also
obtained spectra of its two brightest components, namely ADS~2984A (A component) and SZ~Cam (B component). Since both stars are very close to each other and have very similar photometric and spectroscopic properties, they are often mixed up (see Lorenz, Mayer \& Drechsel 1998, address some catalogues with wrong designations, to avoid further confusion).

The spectra were taken with the fiber spectrograph FIASCO, which is installed at the
Nasmyth focus of the 90\,cm reflector telescope of the University Observatory Jena. The spectrograph exhibits a dispersion of
$\sim$\,0.91\,\AA/pixel and a resolving power of ${R\sim5500}$ at the wavelength of H$\alpha$.
In its actual configuration FIASCO is adjusted to cover the wavelength range from about 6130 up to
7060\,\AA. For a detailed overview of all properties of the spectrograph we refer here to Mugrauer
\& Avila (2009).

Spectra of ADS~2984A and SZ~Cam were taken with FIASCO one after another at similar airmass in two
nights on 2008 October 18 and November 16. Three 2\,min spectra of each star were taken in
October 2008, and two 600\,s spectra for each in November, respectively. For calibration we always
took three dark frames with the same integration time as chosen for spectroscopy. Flat field and
arc-lamp images were taken with the integrated calibration unit of the spectrograph. The reduction
of all data, the extraction of the spectra from the individual images, as well as their wavelength
calibration was done with standard {IRAF} tasks for the reduction of spectroscopic data. The
individual reduced and calibrated spectra of each star were finally averaged, and the continuum was
then normalized to one.

\begin{figure}[t]
  \centering
   \resizebox{0.7\hsize}{!}
{
   \includegraphics[viewport= 85 0 490 372, width=0.48\textwidth]{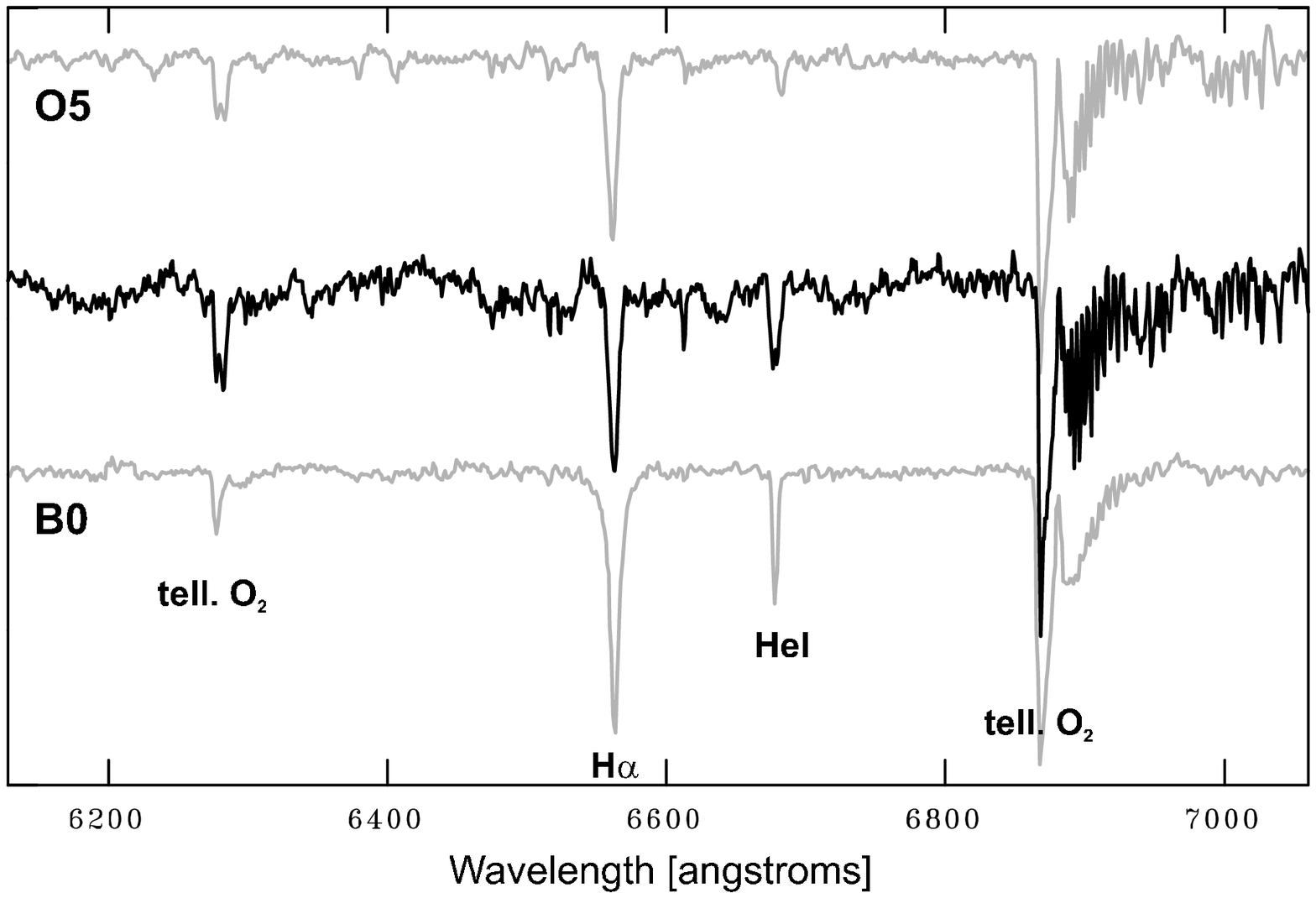}
}
   \caption{Normalized FIASCO spectra of the A component of NGC~1502 (ADS~2984A) from October 2008 (\emph{middle}) and the comparison spectra of early-type dwarfs, taken from Le Borgne et al. (2003).}
              \label{fig1}%
\end{figure}

\begin{figure}[t]
  \centering
   \resizebox{0.7\hsize}{!}
{
   \includegraphics[viewport= 85 0 490 372, width=0.48\textwidth]{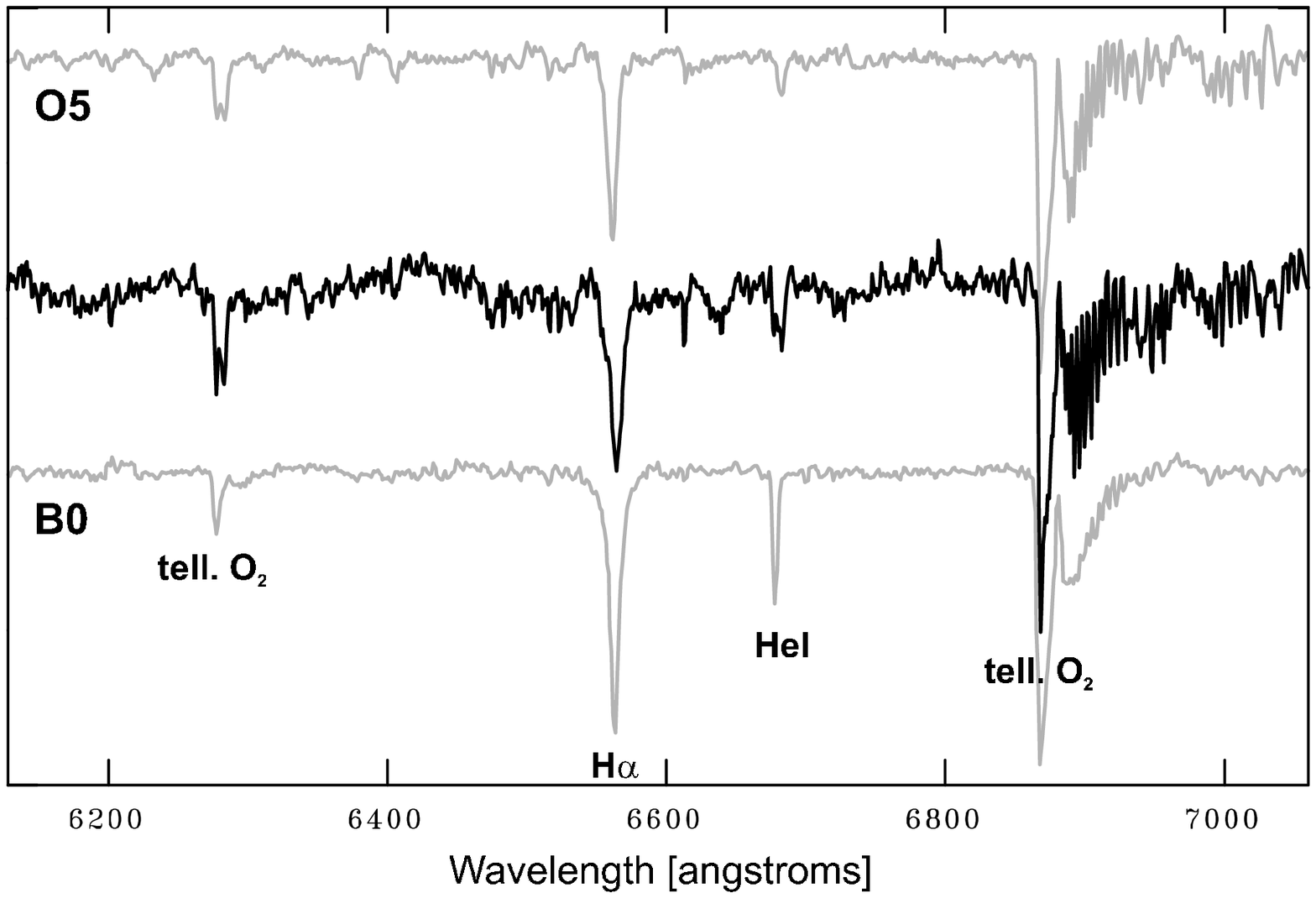}
}
   \caption{Same as in Fig.~\ref{fig1}, but for the B component of NGC~1502 (SZ~Cam).}
              \label{fig2}%
\end{figure}

Figures~\ref{fig1} and \ref{fig2} show the normalized spectra of both components from October 2008. In addition, we
also show normalized spectra of dwarfs, taken from the spectral library of Le Borgne et al. (2003) for comparison.

\begin{figure}
\hskip-10mm
   \resizebox{1.10\hsize}{!}
{
   \includegraphics[viewport=-75 0 490 272, width=0.48\textwidth]{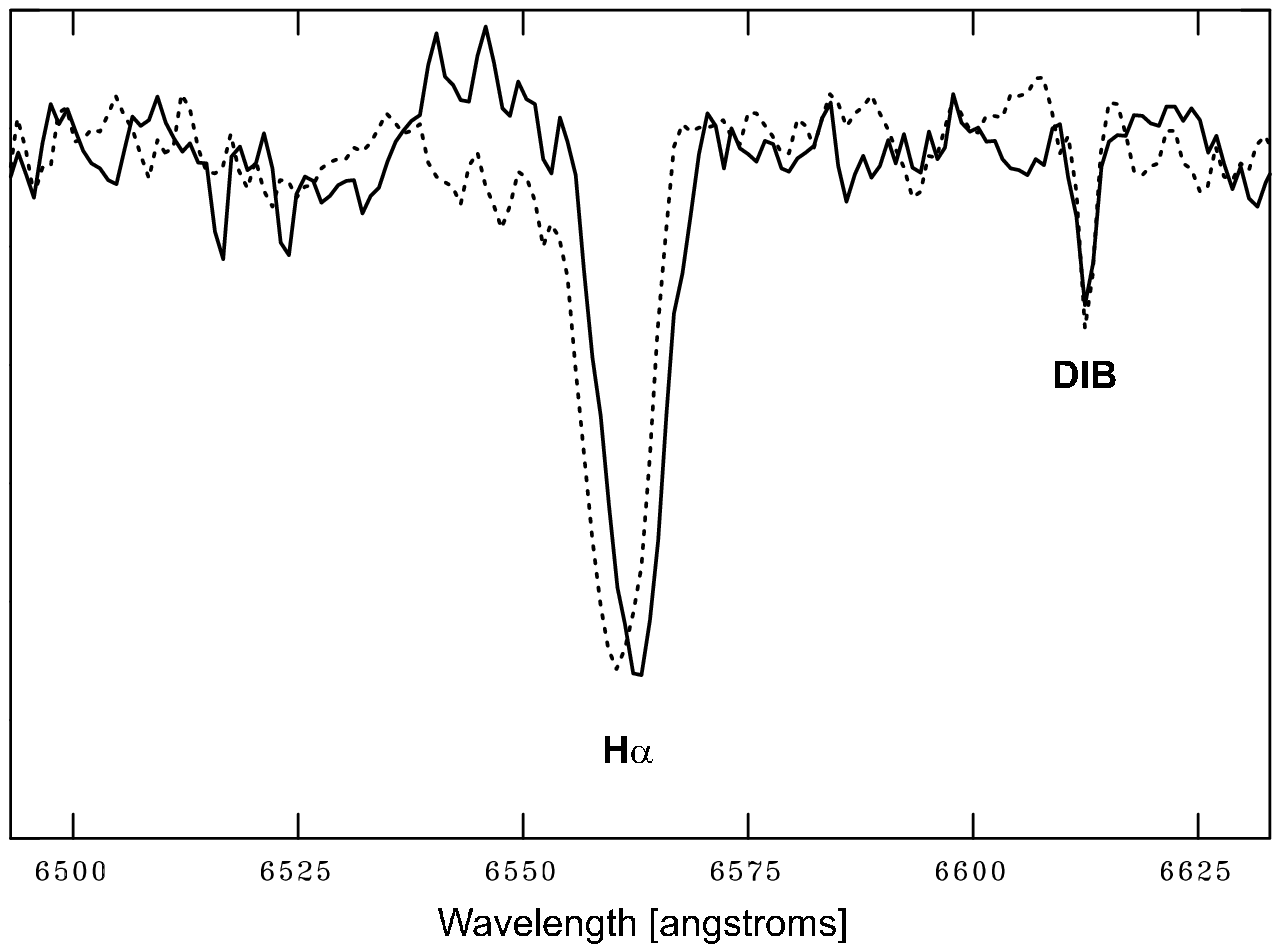}
}
   \caption{The variability of the H$\alpha$ line from the A component of
   NGC~1502 (ADS~2984A) caused by a massive companion, see Lorenz, Mayer \&
   Drechsel (1998). Both spectra are fixed on the telluric oxygen absorption
   features (see Figs. \ref{fig1} and \ref{fig2}) and are constant to the diffuse interstellar band (DIB). Spectra were taken with about one month of epoch difference.}
              \label{fig3}%
\end{figure}
\begin{figure}
\hskip-10mm
   \resizebox{1.1\hsize}{!}
{
   \includegraphics[viewport=-75 0 490 272, width=0.48\textwidth]{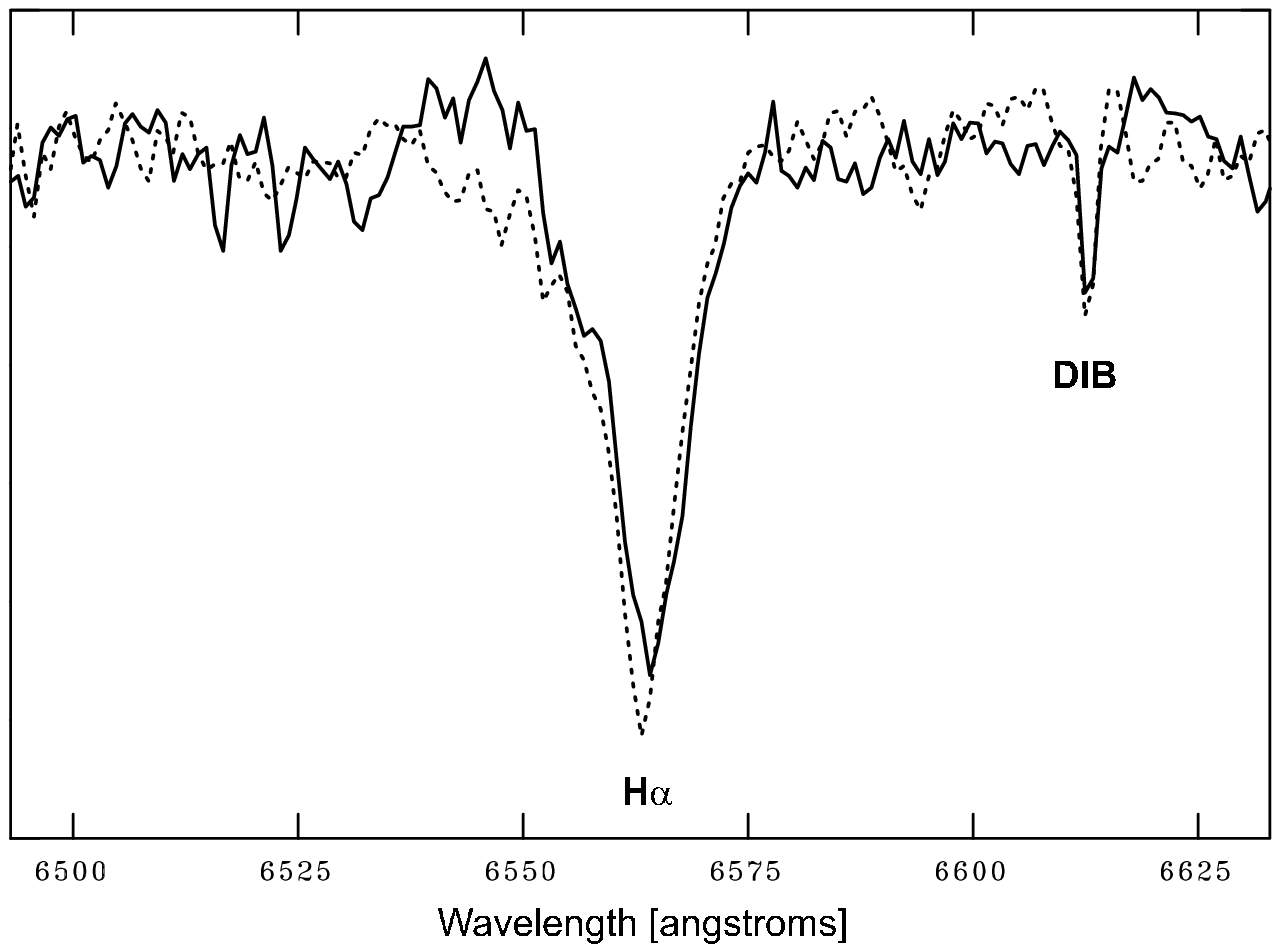}
}
   \caption{Same as in Fig.~\ref{fig3}, but for SZ~Cam. This system is known as double lined spectroscopic binary consisting of a O9\,IV and B0\,V star, see Lorenz, Mayer \& Drechsel (1998).}
              \label{fig4}%
\end{figure}

\begin{figure}[t]
 
   \resizebox{0.90\hsize}{!}
{
   \includegraphics[ width=0.40\textwidth]{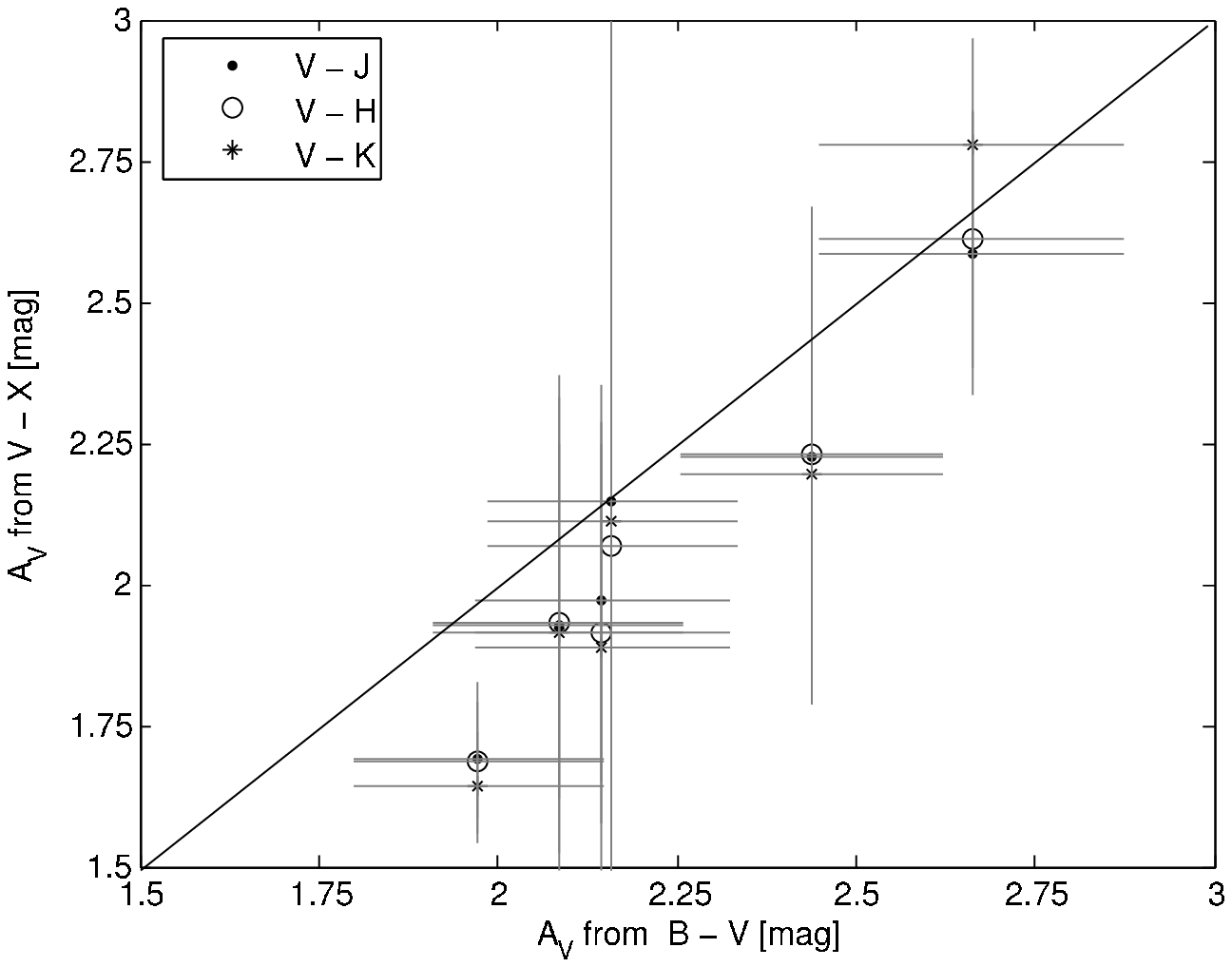}
}
   \caption{$A_{V}$ values obtained from $BV\!JHK$ photometry ($BV$ using CTK and $JHK$ from 2MASS) from the components of the cluster NGC~1502 with known spectral type. Error bars denote one
 sigma.}
              \label{AVnsv}%
    \end{figure}

\begin{figure}[t]

   \resizebox{0.90\hsize}{!}
{
\includegraphics[ width=0.40\textwidth]{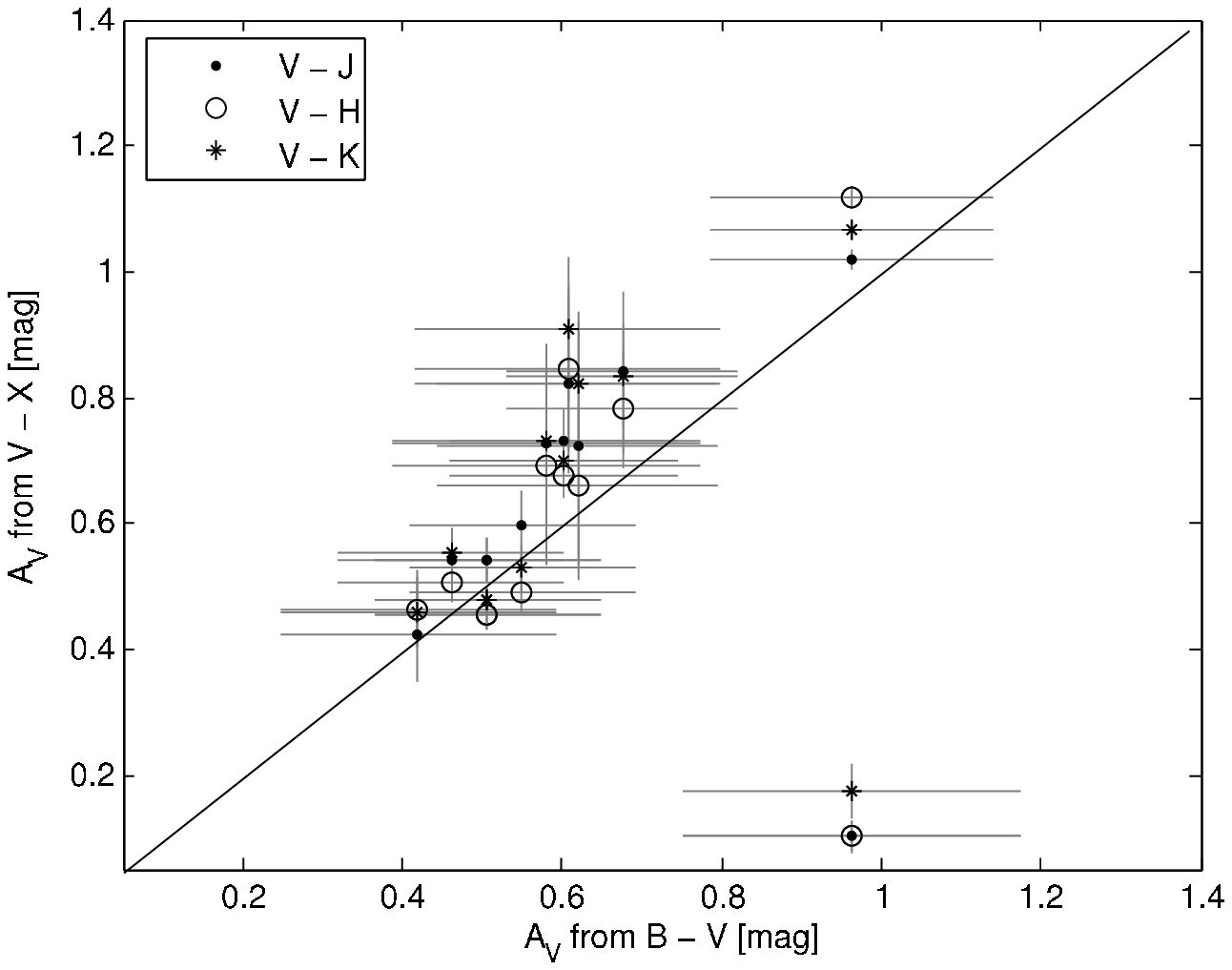}
}
   \caption{$A_{V}$ values obtained from $BV\!JHK$ photometry ($BV$ using CTK and $JHK$ from 2MASS) from the components of the cluster NGC~2169 with known spectral type. The outlier is component C: the $BV$ photometry is affected by the bright components A+B. Error bars denote one sigma.}
              \label{AVhip}%
    \end{figure}

The FIASCO spectra of both components are very similar to each other. Beside the telluric
absorption features of oxygen at 6277\,\AA~and 6867\,\AA, the absorption lines of H$\alpha$
at 6563\,\AA, as well as of He\,{\sc i} at 6678\,\AA~are clearly detected. In the FIASCO spectra the
H$\alpha$ and He\,{\sc i} line are both fainter than in the B0V comparison spectrum, which indicates a
spectral type of both components earlier than B0. The detected H$\alpha$ lines fit well with
that of the O5V comparison spectrum, while the He\,{\sc i} lines are both stronger, consistent with O-type
stars with a spectral type later than O5. Hence, from our FIASCO spectroscopy we can conclude that
the spectral type of ADS~2984A and SZ~Cam should range between O5 and B0.

Details of all FIASCO spectra, centered around the H$\alpha$-line, are shown in Fig.~\ref{fig3} for
ADS~2984A, and in Fig.~\ref{fig4} for SZ~Cam, respectively. Beside the strong hydrogen line our
FIASCO spectra also show the absorption line of diffuse interstellar bands (DIB) at 6614\,\AA. We
determined the center of the DIB and H$\alpha$ line in all spectra using Gaussian line fitting
with {ESO-MIDAS}. Between October and November 2008 we do not find a significant shift of
the DIB line in the two spectra of each stars. In contrast, in the spectra of both stars a
significant shift of the H$\alpha$ line-center is detected between both observing epochs. We find
a shift of the H$\alpha$ line-center of $-1.99\pm0.14$\,\AA~in the case of ADS~2984A and
$-0.87\pm0.13$\,\AA~for SZ~Cam, respectively. The shift of the H$\alpha$ line between both
observing epochs induced by the change of the heliocentric velocity is expected to be only
$-0.21$\,\AA. Hence, the detected line shifts in the spectra of ADS~2984A and SZ~Cam clearly confirm
that they are both spectroscopic multiple systems. This spectroscopic investigation of well known binaries shows that the conditions at the University Observatory Jena are sufficient to detect binaries in comparable OB~clusters in further investigations. 
ADS2984A, as well as SZ Cam are both known as double lined spectroscopic
binaries (Lorenz,Mayer \& Drechsel 1998). The detection of the double
lined spectroscopic nature of massive binaries like ADS2984A, or SZ Cam is
challenging with FIASCO, discernible only as a small asymmetry of their
line profiles (see e.g. ~\ref{fig4}). However, with multi-epoch radial velocity
measurements the multiplicity of these stars can be clearly revealed with
FIASCO.


\section{Results}

\subsection{Magnitudes, colours, absorptions, and spectral types}

\begin{table*}[t]
\caption{Medians of the magnitudes of the components of NGC~1502. The errors are the mean error of the single images. The $A_{V}$ values of each member are calculated from the median of the $A_{V}$
values obtained from the different $BV\!JHK$ colours, the error denotes the standard deviation. The $BV\!RI$ magnitudes listed here are obtained with the CTK, while $JHK$ are 2MASS magnitudes.}
\label{nsvtab}
\begin{tabular}{lcccccc}
\hline\noalign{\smallskip}
Component & $B$	& $V$ & $R$ &	$I$ & $A_{V}$($BV\!JHK$) \\
 					& [mag]	& [mag]	& [mag] & [mag] & [mag] \\[1.5pt]
\hline\noalign{\smallskip}

A &	7.349	$\pm$ 0.027	& 6.880	$\pm$ 0.018	& 6.563	$\pm$ 0.018	& 6.271	$\pm$ 0.022 &  1.93 $\pm$ 0.08\\
B &	7.338	$\pm$ 0.027	& 6.910	$\pm$ 0.018	& 6.635	$\pm$ 0.018	& 6.387	$\pm$ 0.022 &  2.13 $\pm$ 0.04\\
C+G &	7.351	$\pm$ 0.027	& 6.880	$\pm$ 0.018	& 6.838	$\pm$ 0.018	& 6.954	$\pm$ 0.022 &    -\\
D &	9.890	$\pm$ 0.027	& 9.401	$\pm$ 0.019	& 9.112	$\pm$ 0.018	& 8.820	$\pm$ 0.022 &  1.94 $\pm$ 0.00\\
E &	11.571	$\pm$ 0.033\enspace	& 11.003 $\pm$ 0.026\enspace	& 10.572 $\pm$ 0.022\enspace	& 10.163 $\pm$ 0.026\enspace &       -\\
F &	7.352	$\pm$ 0.027	& 6.880	$\pm$ 0.018	& -			& - 		    &      -\\
G+C &	7.351	$\pm$ 0.027	& 6.880	$\pm$ 0.018	& 6.838	$\pm$ 0.018	& 6.954	$\pm$ 0.022 &        -\\
H+I &	9.734	$\pm$ 0.027	& 9.187	$\pm$ 0.019	& 8.854	$\pm$ 0.018	& 8.524	$\pm$ 0.022 &      -\\
J &	10.273	$\pm$ 0.027\enspace	& 9.732	$\pm$ 0.019	& 9.406	$\pm$ 0.018	& 9.083	$\pm$ 0.022 &  1.69 $\pm$ 0.15\\
K+L &	10.316	$\pm$ 0.027\enspace	& 9.771	$\pm$ 0.019	& 9.437	$\pm$ 0.018	& 9.090	$\pm$ 0.022 &     -\\
M &	11.616	$\pm$ 0.033\enspace	& 11.008 $\pm$ 0.026\enspace	& 11.051 $\pm$ 0.027\enspace	& 10.589 $\pm$ 0.031\enspace &      -\\
N &	10.159	$\pm$ 0.027\enspace	& 9.589	$\pm$ 0.019	& 9.242	$\pm$ 0.018	& 8.894	$\pm$ 0.022 &  2.23 $\pm$ 0.11\\
O &	11.487	$\pm$ 0.031\enspace	& 10.843 $\pm$ 0.023\enspace	& 10.462 $\pm$ 0.020\enspace	& 10.031 $\pm$ 0.024\enspace &  2.64 $\pm$ 0.09\\
P &	12.091	$\pm$ 0.039\enspace	& 11.500 $\pm$ 0.033\enspace	& 11.132 $\pm$ 0.027\enspace	& 10.722 $\pm$ 0.031\enspace &      -\\[1.5pt]
\hline
\end{tabular}
\end{table*}

\begin{table*}[t]
\caption{Same as in Table \ref{nsvtab}, but with the cluster NGC~2169. The P component is not listed in 2MASS, i.e. we derived the $A_{V}$ value from ${B-V}$ measured from this work only.}
\label{hiptab}
\begin{tabular}{lcccccc}
\hline\noalign{\smallskip}
Component & $B$	& $V$ & $R$ &	$I$ &  $A_{V}$($BV\!JHK$)\\ 
 					& [mag]	& [mag] & [mag] & [mag] &  [mag] \\[1.5pt]
\hline\noalign{\smallskip}

A+B &	6.853	$\pm$ 0.022 & 6.906	$\pm$ 0.022 & 6.962	$\pm$ 0.031 & 6.976	$\pm$ 0.028 & 0.49 $\pm$ 0.04\\
C &	11.198	$\pm$ 0.035\enspace & 10.948	$\pm$ 0.029\enspace & 11.093	$\pm$ 0.042\enspace & 10.703	$\pm$ 0.037\enspace & 0.14 $\pm$ 0.42\\
D &	8.562	$\pm$ 0.022 & 8.599	$\pm$ 0.022 &  8.649	$\pm$ 0.031 & 8.638	$\pm$ 0.028 & 0.69 $\pm$ 0.05\\
E &	9.173	$\pm$ 0.022 & 9.134	$\pm$ 0.022 & 9.133	$\pm$ 0.031 & 9.069	$\pm$ 0.028 & 0.81 $\pm$ 0.08\\
F &	11.028	$\pm$ 0.030\enspace & 9.891	$\pm$ 0.022 & 9.336	$\pm$ 0.031 & 8.761	$\pm$ 0.028 & 1.04 $\pm$ 0.07\\
G &	11.923	$\pm$ 0.052\enspace & 11.836	$\pm$ 0.046\enspace & 11.830	$\pm$ 0.065\enspace & 11.710	$\pm$ 0.071\enspace & -  \\
H &	11.034	$\pm$ 0.034\enspace & 10.896	$\pm$ 0.027\enspace & 10.861	$\pm$ 0.038\enspace & 10.703	$\pm$ 0.037\enspace & -  \\
I &	8.713	$\pm$ 0.022 & 8.745	$\pm$ 0.022 &  8.795	$\pm$ 0.031 & 8.779	$\pm$ 0.028 & 0.52 $\pm$ 0.04\\
J &	10.807	$\pm$ 0.028\enspace & 10.757	$\pm$ 0.026\enspace & 10.760	$\pm$ 0.037\enspace & 10.682	$\pm$ 0.036\enspace & 0.69 $\pm$ 0.09\\			
L &	10.851	$\pm$ 0.028\enspace & 10.761	$\pm$ 0.026\enspace & 10.739	$\pm$ 0.037\enspace & 10.598	$\pm$ 0.035\enspace & 0.44 $\pm$ 0.02\\
M &	11.148	$\pm$ 0.032\enspace & 11.025	$\pm$ 0.029\enspace & 10.959	$\pm$ 0.040\enspace & 10.853	$\pm$ 0.039\enspace & 0.71 $\pm$ 0.07\\
N &	8.058	$\pm$ 0.022 & 8.061	$\pm$ 0.022 & 8.073	$\pm$ 0.031 & 8.071	$\pm$ 0.028 & 0.54 $\pm$ 0.04\\
O &	11.243	$\pm$ 0.030\enspace & 11.111	$\pm$ 0.029\enspace & 11.083	$\pm$ 0.041\enspace & 10.915	$\pm$ 0.039\enspace & 0.83 $\pm$ 0.13\\
P &	10.579	$\pm$ 0.028\enspace & 10.531	$\pm$ 0.025\enspace & 10.542	$\pm$ 0.035\enspace & 10.449	$\pm$ 0.034\enspace & 0.35 $\pm$ 0.16 \\[1.5pt]
\hline
\end{tabular}
\end{table*}

Starting from absolute $BV\!JHK$ magnitudes and colours we have enough information to calculate the $A_{V}$ values for given spectral types (as listed in Simbad) of the members. 

The (constant) ratios of $A_{V}$ to the absorptions of different bands $X$, i.e.
$A_{V}$/$A_{X}$, are given in the models of Rieke \& Lebofsky (1979), Savage \&
Bolton (1979), and Cardelli, Clayton \& Mathis (1989), whereas we used the
medians of the different values from the different authors of each ratio. The stellar
atmosphere models of Bessell, Castelli \& Plez (1998) provide for each
effective temperature the model colours $B-V$, $V-R$ and $V-I$, and bolometric
corrections up to 50\,000~K, and Kenyon \& Hartmann (1995) list $B-V$, $V-J$, $V-H$ and
$V-K$, and bolometric corrections for B0 and later. Moreover, Schmidt-Kaler (1982) lists bolometric corrections for stars earlier than B0, which completes our model data. 

With given model colours and
the measured data we estimated the absorption by calculating the $A_{V}$ values
in each band. The formal errors of the $A_{V}$ values are derived by the errors
of the measured magnitudes and colours. The $A_{V}$ values from $B-V$, $V-J$, $V-H$,
and $V-K$ for both clusters are shown in Figs. \ref{AVnsv} and \ref{AVhip} and almost all of them
follow a one-to-one relation within their one sigma error bars, although there is a tendency for a systematic deviation from this relation for both clusters. This is maybe caused due to the different ratios of $A_{V}$ to the absorptions of different bands in different directions, which contradicts the constant ratios assumed by the models, mentioned before. We calculated the medians of the $A_{V}$ values
and their standard deviation and list them in Tables~\ref{nsvtab} and \ref{hiptab}. 

While the median value of absorption of the unresolved binary SZ~Cam is $A_{V} = 2.13\pm0.04$~mag from our calculations (and 2.16~$\pm$~0.18~mag from $B-V$ only), Crawford (1994) lists 2.32~mag from $B-V$. Lorenz, Mayer \& Drechsel (1998) give a total visual magnitude of ${V=6.92}$~mag for SZ~Cam (6.910~$\pm$~0.018 from own photometry) and ${V=7.7}$ and 8.63~mag for the resolved components. We use these values for the mass estimation in the next section.

\begin{table*}[t]
\vskip-1mm
\caption{Spectral types as listed in Simbad (for second line from  Lorenz, Mayer \& Drechsel 1998) and masses for NGC~1502. The mass determination is based on the luminosity and, hence the
assumed distance (here 836~pc and 1050~pc, see Lorenz, Mayer \& Drechsel 1998). The errors for the luminosities are derived from the two distances, while the mean value is given. The mass values are obtained using the evolutionary models from Bertelli~et~al. (1994) (B), Claret (2004) (C) and Schaller~et~al. (1992) (S).}
\label{nsvtabmass}
\begin{tabular}{llc|cccc|cccc}
\hline
Component & Sp. Type &$\log L_{\rm bol}$ &\multicolumn{4}{c}{Mass for 836~pc} & \multicolumn{4}{c}{Mass for 1050~pc} \\ 
 					& & & B & C & S & Mean $\pm$ Std. & B & C & S &Mean $\pm$ Std.\\
 					& &[L$_{\odot}]$ & \multicolumn{4}{c|}{[M$_{\odot}$]}& \multicolumn{4}{c}{[M$_{\odot}$]}\\[1.5pt] 					
\hline
 & & & & & & & & & &  \\[-8pt]

A &	B0II 	&  $4.71^{+0.09}_{-0.11}$  &  13.0 &  15.6  & 14.8  & 14.5  $\pm$  1.3	 &	14.9 &  15.5 &  14.7 &  15.1 $\pm$  0.4	\\[3pt]
B &	O9 IV &  $4.86^{+0.09}_{-0.11}$     &  19.0 &  19.8  & 19.8  & 19.5  $\pm$  0.5  &	22.1 & \enspace 19.8\footnotemark[1]  &  22.2 &  21.4 $\pm$  1.4	\\[3pt]
  &	+ B0V &  $4.49^{+0.09}_{-0.11}$     &  14.7 &  15.8  & 15.0  & 15.2  $\pm$  0.6  &	17.1 &  17.7 &  14.8 &  16.6 $\pm$  1.5	\\[3pt]
D &	B2V   &  $3.71^{+0.09}_{-0.11}$     & \enspace 8.5  &\enspace  7.9   &\enspace 9.0   & \enspace8.5   $\pm$  0.5  &\enspace	9.0  &  10.0 &  \enspace9.0  &\enspace  9.3 $\pm$  0.5	\\[3pt]
J &	B8V   &  $2.89^{+0.09}_{-0.11}$     & \enspace 4.2  & \enspace 4.0   &\enspace 4.8   &\enspace 4.3   $\pm$  0.4  &\enspace	4.6  & \enspace 5.0    & \enspace 4.3  &\enspace  4.6 $\pm$  0.4	\\[3pt]
N &	B1V   &  $3.87^{+0.09}_{-0.11}$     & \enspace 9.8  &  10.0  &\enspace 9.0   &\enspace 9.6   $\pm$  0.5  &	11.0 &  10.0 & \enspace 9.0  &  10.0 $\pm$  1.0	\\[3pt]
O &	B1.5V &  $3.53^{+0.09}_{-0.11}$     & \enspace 7.9  & \enspace 7.9   &\enspace 7.0   &\enspace 7.6   $\pm$  0.5  &\enspace	8.8  & \enspace 7.9  & \enspace 9.0  & \enspace 8.6 $\pm$  0.6	\\[2.5pt]
\hline
\end{tabular}
\\
$^1$ for Z=0.04, because the model for solar metallicity produces 6.6 M$_{\odot}$, which is obviously inconsistent with the other masses.
\end{table*}


\begin{table*}[t]
\vskip-1mm
\caption{Same as in Table \ref{nsvtabmass}, but with the cluster NGC~2169. Masses are calculated for the distance listed in Hipparcos (376~pc) and Roth \& Warman (1979) (639~pc).
The masses for the smaller distance seem to be too small for the corresponding spectral types, i.e. the 639~pc is more realistic. Although in this case the masses of the components C, D, M, O and P are too small (P has no 2MASS magnitudes). These stars probably do not belong to this cluster and may have a larger distance.}
\label{hiptabmass}
\begin{tabular}{llc|cccc|cccc}
\hline
Component & Sp. Type &$\log L_{\rm bol}$ &\multicolumn{4}{c}{Mass for 376~pc} & \multicolumn{4}{c}{Mass for 639~pc} \\ 
 					& & & B & C & S & Mean $\pm$ Std. & B & C & S & Mean $\pm$ Std.\\
 					& &[L$_{\odot}]$ & \multicolumn{4}{c|}{[M$_{\odot}$]}& \multicolumn{4}{c}{[M$_{\odot}$]}\\[1.5pt] 					
\hline
 & & & & & & & & & &  \\[-8pt]

A+B &	B2V &  $3.61^{+0.17}_{-0.29}$  & 				7.4 &  7.9 & 7.0  &  7.5  $\pm$  0.5  & 9.1   &  10.0 &   9.0  &   9.4 $\pm$ 0.6\\[3pt]
C &	B9V &  $1.16^{+0.17}_{-0.29}$&				1.6 &  1.6 & 1.6  &  1.6  $\pm$  0\enspace\ \    & 2.1   &  2.0  &   1.8  &   2.0 $\pm$ 0.1\\[3pt]
D &	B1V &  $3.13^{+0.17}_{-0.29}$& 				5.3 &  5.0 & 4.6  &  5.0  $\pm$  0.3  & 7.1   &  6.3  &   7.0  &   6.8 $\pm$ 0.4\\[3pt]
E &	B3V &  $2.70^{+0.17}_{-0.29}$& 				4.1 &  4.0 & 3.7  &  3.9  $\pm$  0.2  & 5.6   &  5.0  &   4.7  &   5.1 $\pm$ 0.4\\[3pt]
F &	G8V &  $1.82^{+0.17}_{-0.29}$& 				2.4 &  2.3 & 1.6  &  2.1  $\pm$  0.4  & 3.0   &  3.2  &   2.5  &   2.9 $\pm$ 0.3\\[3pt]
I &	B3V &  $2.75^{+0.17}_{-0.29}$& 				4.2 &  4.0 & 3.8  &  4.0  $\pm$  0.2  & 5.6   &  5.0  &   4.7  &   5.1 $\pm$ 0.5\\[3pt]
J &	B5  &  $1.79^{+0.17}_{-0.29}$& 				2.3 &  2.2 & 2.2  &  2.2  $\pm$  0.1  & 3.1   &  3.2  &   2.8  &   3.0 $\pm$ 0.2\\[3pt]
L &	A0  &  $1.28^{+0.17}_{-0.29}$& 				1.7 &  1.8 & 1.6  &  1.7  $\pm$  0.1  & 2.3   &  2.0  &   1.8  &   2.0 $\pm$ 0.2\\[3pt]
M &	B9.5&  $1.36^{+0.17}_{-0.29}$& 				1.8 &  1.8 & 1.6  &  1.7  $\pm$  0.1  & 2.4   &  2.5  &   2.5  &   2.5 $\pm$ 0.1\\[3pt]
N &   B2.5V &  $4.71^{+0.17}_{-0.29}$& 				5.0 &  5.0 & 4.8  &  4.9  $\pm$  0.1  & 6.4   &  6.3  &   6.2  &   6.3 $\pm$ 0.1\\[3pt]
O &	B9V &  $1.37^{+0.17}_{-0.29}$& 				1.8 &  1.8 & 1.8  &  1.8  $\pm$  0\enspace\ \     & 2.4   &  2.5  &   2.5  &   2.5 $\pm$ 0.1\\[3pt]
P &	B9V & $0.96^{+0.23}_{-0.23}$	&			1.5   &	1.5  &  1.4  &   1.5 $\pm$ 0.1    &  2.0    &  2.0    &   1.9   &  2.0  $\pm$ 0.1\\[2pt]

\hline
\end{tabular}
\end{table*}

\subsection{Mass estimation}

Knowing $V$-band magnitudes, bolometric corrections, and absorptions one can calculate the luminosity for a given distance and from spectral type we derived the temperature. With temperature and luminosity we can now estimate the mass of the star using model grids of stellar evolution by putting the star into the HR-Diagram. Assuming solar  metallicity we used the models from Bertelli~et~al. (1994) (up to 35~M$_{\odot}$), Schaller~et~al. (1992) and Claret (2004) (both up to 125~M$_{\odot}$). We interpolated linearly between the evolutionary tracks. Each model takes mass loss into account.

Since the luminosity is very sensitive to the parallax, we calculated the mass
for 836~pc and 1050~pc, see Lorenz, Mayer \& Drechsel (1998) for NGC~1502 and for
376~pc (Hipparcos) and 639~pc (Roth \& Warman 1979) for  NGC~2169.
For some components the mass value is obviously not consistent to the spectral
type (see Tables~\ref{nsvtabmass} and \ref{hiptabmass}). Likely these stars do
not belong to the cluster and are more distant (C, D, M, O and P in NGC~2169).
Unfortunately almost none of these stars have trigonometric parallaxes in
Hipparcos, so that we cannot prove this claim directly. Only the A and B
components for both clusters and I and N for NGC~2169 have Hipparcos parallaxes
(the parallax value for the D component in NGC~2169 has a negative value), but
the latter ones have reasonable masses within their errors and the parallaxes
correspond to 600\,--\,660~pc for both. As discussed in the first section, the distances of both clusters are rarely known and have a strong influence on the resulting masses obtained from the luminosities. The WEBDA data base gives 1052~pc for NGC~2169, which differs much from these distances used in Tables~\ref{hiptabmass} and would increase the masses significantly. This has to be considered if ages and residual life times (derived from the masses) are used for further investigations as suggested in Hohle, Neuh\"auser \& Tetzlaff (2008). 

From radial velocity curves of SZ~Cam it is known that both components have 15
and 11~M$_{\odot}$ (Lorenz, Mayer \& Drechsel 1998), i.e. 26~M$_{\odot}$
together. The mass estimation from the luminosity of this component, which is
unresolved in the CTK, results in 19.5\,$\pm$\,0.5 and 15.2\,$\pm$\,0.6~M$_{\odot}$
or 21.4\,$\pm$\,1.4 and 16.6\,$\pm$\,1.5~M$_{\odot}$ for a distance of 836~pc or
1050~pc, respectively (see Table~\ref{nsvtabmass}, component B), using the visual magnitudes given in Lorenz, Mayer \& Drechsel (1998).

\section{Conclusion}

We performed absolute $BV\!RI$ photometry at the University Observatory Jena with the Cassegrain-Telescop-Kamera  (CTK) of sixteen members (as listed in the CCDM) of the OB clusters NGC~1502 and NGC~2169 including those stars which were not measured in Hipparcos. We calculated the absorption due to the interstellar medium for the components with known spectral type using colour models. Moreover, we estimated the masses of these stars, using $BV\!JHK$ photometry for different distances of each cluster. 
The obtained values for masses and absorptions are reliable for the inner components of the cluster, but for the outer components the mass tends to too low values. These stars likely do not belong to the clusters physically and have a larger distance. The mass estimation based on the luminosities of the stars strongly depends on the assumed distance.

Since NGC~1502 is relatively well investigated and our methods work in general, NGC~2169 should be an object for further investigations for spectroscopy. 
As described in Sect. 3, we are able to detect the multiple character of individual cluster components with FIASCO at the 90~cm telescope of the University Observatory Jena, which was done for the first time at this Observatory. Thus, a detailed spectroscopic monitoring of NGC~2169 seems promising for further binary detections.

\acknowledgements
RN acknowledges general support from the German National Science
Foundation (Deutsche Forschungsgemeinschaft, DFG) in grants NE 515/13-1,
13-2, and 23-1,
SR and MV acknowledge support from the EU in the FP6 MC ToK project
MTKD-CT-2006-042514,
TOBS acknowledges support from the Evangelisches Studienwerk e.V.
Villigst,
TE, NT and MMH acknowledge partial support from DFG in the SFB/TR-7 Gravitation
Wave Astronomy and MM acknowledges support from the government of Syria.
Furthermore we aknowledge the 2MASS point source catalogue.





\end{document}